\documentclass[a4paper]{article}

\usepackage[pages=all, color=black, position={current page.south}, placement=bottom, scale=1, opacity=1, vshift=5mm]{background}
\SetBgContents{
	\tt Bhowmik et al. 2023: Manuscript submitted to Journal
}      % copyright

\usepackage[margin=1in]{geometry} % full-width

\usepackage{amsmath}
\usepackage{amsthm}
\usepackage{amssymb}

\usepackage[utf8]{inputenc}
\usepackage{hyperref}
\hypersetup{
	unicode,
	pdfauthor={Author One, Author Two, Author Three},
	pdftitle={A simple article template},
	pdfsubject={A simple article template},
	pdfkeywords={article, template, simple},
	pdfproducer={LaTeX},
	pdfcreator={pdflatex}
}

\usepackage[sort&compress,numbers,square]{natbib}
\theoremstyle{plain}

\theoremstyle{definition}

\usepackage{graphicx, color}
\graphicspath{{fig/}}

\usepackage{algorithm, algpseudocode} % use algorithm and algorithmicx for typesetting algorithms
\usepackage{mathrsfs} % for \mathscr command

\usepackage{lipsum}

\title{Eulerian-Lagrangian particle-based model for diffusional growth for the better parameterization of ISM clouds: A road map for improving climate model through small-scale model using observations}
\author{Moumita Bhowmik$^1$\thanks{Corresponding Author: moumita.bhowmik@tropmet.res.in} \and Anupam Hazra$^1$\thanks{Corresponding Author: hazra@tropmet.res.in} \and Suryachandra A. Rao$^1$ \and Lian-Ping Wang$^2$}

\date{
	$^1$Indian Institute of Tropical Meteorology, Ministry of Earth Sciences, Pune, India \\ %
	$^2$Center for Complex Flows and Soft Matter Research, Department of Mechanics and Aerospace Engineering, Southern University of Science and Technology, China \\ [2ex]%
}

\begin{document}
	\maketitle
	
	\begin{abstract}
		The quantitative prediction of the intensity of rainfall events (light or heavy) has remained a challenge in Numerical Weather Prediction (NWP) models. For the first time the mean coefficient of diffusional growth rates ($c_m$) are calculated using an Eulerian-Lagrangian particle-based small-scale model on in situ airborne measurement data of Cloud Aerosol Interaction and Precipitation Enhancement Experiment (CAIPEEX) during monsoon over Indian sub-continent. The results show that $c_m$ varies in the range of $\sim$  0.25$\times 10^{-3}$– $1.5\times 10^{-3}$(cm/s). The generic problem of the overestimation of light rain in NWP models might be related with the choice of $c_m$ in the model. It is also shown from DNS experiment using Eulerian-Lagrangian particle-based small-scale model that the relative dispersion ($\epsilon$) is constrained with average values in the range of $\sim$ 0.2–0.37 ($\sim$ 0.1–0.26) in less humid (more humid) conditions. This is in agreement with in situ airborne observation ($\epsilon$ ~$\sim$  0.36) and previous study over Indian sub-continent. The linear relationship between relative dispersion ($\epsilon$) and cloud droplet number concentration (NC) is obtained from this study using CAIPEEX observation over Indian subcontinent. The dispersion based “autoconversion” scheme for Indian region must be useful for the Indian summer monsoon precipitation calculation in the general circulation model. The present study also provide valuable guidance for the parameterization of effective radius, important for radiation scheme.

	\end{abstract}

	%\tableofcontents
	
	\section{Introduction}
Accurate quantitative prediction of Indian summer monsoon (ISM) rainfall using climate model is important for weather forecasting. However, forecasting of precipitation is challenging because the precipitation process depends not only on the synoptic situations but also on the cloud microphysical processes \cite{Shr13}. Most of the climate models are insufficient of including microphysical processes explicitly and thus, rely on cloud microphysics parameterization, which is essential for the simulation of organization of mesoscale systems \cite{Haz17,Haz20,Mud18, Mud22, Dutt22a, Dutt22b}. Therefore, understanding and accurately representing physical processes involved in the formation or growth of clouds and rain droplets at “cloud microscale” is important \cite{Grab19}. With increasing computational resources, many sophisticated approaches for accurate modelling of microphysical schemes in climate model are proposed by past researchers \cite{Grab99, Grab01, Rand03, KhaRan01, Bod19}. However, application of these approaches in large-scale models brings significant challenges which are difficult to overcome. Hence, \citet{Grab19} and \citet{Mor20} have advocated alternative approaches such as small-scale cloud modelling and observational advances from laboratory experiments, respectively, to resolve cloud microphysics for large-scale models. However, scaling of a real cloud in laboratory experiments such as the Pi chamber \cite{ChanSaw17, ChanSaw18, ChanCan18, ChanBen16, Des18} is very challenging and need a sustained support for laboratory facilities \cite{Thom19, Mor20}. 
\par
The small-scale cloud model is gaining popularity as it analyses the growth of individual cloud droplets and their rate of formation of warm cloud precipitation at sub-centimeter scales \cite{Grab19}. It is believed that direct numerical simulation (DNS) is the most complete Eulerian-Lagrangian particle-based model that represents turbulent cloud at Kolmogorov length scale by considering all cloud particles individually \cite{And04, And06, Wang09, BurBre07, Kor13, GraWang13}. However, a handful of DNS studies are available to date that analyses droplets growth in turbulent adiabatic ascending domain to explain the growth of cloud droplet size distribution (DSD) \cite{Goh16, Chen18, SaiGot18, Chen20, Chen21}. Though in a vertically oscillating parcel model, it is observed that the curvature effect and the solute effect are essential for DSD broadening \cite{Sri91, Kor95,Yan18}, but using DNS, there are only a few paper \cite{Chen20, Chen21} included these effects to analyse the cloud DSD broadening. \citet{Chen20} developed a hybrid approached, called as “parcel-DNS”, to simulate the relative importance of turbulence and aerosol effects on DSD broadening during the early stage of cloud and rain formation. Their findings indicate that autoconversion rate, defined as the mass transfer rate from droplets smaller than 30 micrometer radii to droplets larger than 30 micrometer radii, is co-related to turbulence intensity as well as shape of the DSD and hence, the traditional autoconversion parameterization, such as Kessler-type or Sundqvist-type autoconversion, should consider the shape/size of DSD along with liquid water content (LWC) and mean droplet radius. Their results also show the importance of consideration of turbulence dependent relative-dispersion parameter. Moreover, \citet{Chen21} concluded that to investigate the microphysical response to seeding giant aerosol particles along with supersaturation fluctuation, DNS is an effective numerical approach. Their studies illustrate that seeding outcome is a joint effect of size, chemical composition and number concentration of seeding aerosols and hence, it is important to include solute term in droplet growth equation.
\par
This study investigates the diffusional growth of Indian summer monsoon (ISM) cloud by DNS using air borne study, namely, Cloud Aerosol Interaction and Precipitation Enhancement Experiments (CAIPEEX) conducted by IITM. In the early stage of cloud development, the activated droplets further grow efficiently by the diffusional growth of water vapors \cite{PruKle10}. The growth of cloud droplets depends on the available moisture and temperature gradient at their immediate vicinity \cite{GraWang13}. As a result, an individual droplet expects to experience different supersaturation and thus, their respective diffusional growth rate must vary. Eventually, it plays a vital role on the rate of formation of warm cloud precipitation in cumulus clouds. Apart from supersaturation fluctuation, to accelerate the diffusional growth of cloud droplets, various other hypotheses including turbulent mixing of clear-cloudy air and atmospheric aerosol-cloud interaction, have also been suggested (see the review articles by \citet{BeaOch93,GraWang13} and there references). 
\par
After diffusional growth, collisional growth is another most important process that controls the formation of warm rain \cite{Chen16,Chen18a,Chen18}. A droplet that manages to grow to a diameter of about 20 $\mu m$ will start to grow by collision and coalescence to form rain drop \cite{PruKle10}. It is important to mention here that, DNS not only provides insightful information about the probability distribution of individual droplets at different heights, total number concentration, and size of droplets to the modelling community, but also the coefficient of diffusional growth rate and relative dispersion of different ISM clouds can be obtained with the help of observational data. These coefficients or the values of parameters  might be useful in climate model to improve realistic simulation of light and moderate rain during ISM \cite{Fio11, Mor20}. 
\par
To the best of our knowledge, this is the first time the range of the coefficient of diffusional growth rate and dispersion based ‘autoconversion’ of ISM cloud is being obtained using upward rising DNS. The aircraft observations of ISM convective clouds distribution are used as initial distributions to step up the simulations. These aircraft observations of clouds were made during Cloud Aerosol Interaction and Precipitation Enhancement Experiment (CAIPEEX) \cite{Kul12} conducted over the Indian sub-continent. In this study, we investigate the diffusional growth of cloud droplets in two different atmospheric conditions (i.e., less and more humid). The role of different sizes of dry aerosol on the growth process is also investigated. 
\par
This study attempts to address the following questions:\\
  1) what is the role of dry aerosol size on diffusion growth of ISM cloud droplets and the relative dispersion?\\
  2) what is the role of relative humidity on diffusion growth and broadening spectra of ISM shallow (less humid) and convective (more humid) clouds?\\
  3) what is the importance of relative dispersion for revisiting modified 'autoconversion' parameterization for climate model?\\
  4) how to parameterize effective radius from DNS experiments for radiation scheme in climate model?
\par
The present paper is organized as follows. Section 2 introduces the description of model, initial setup and design of experiments for the DNS model. The results from sensitivity studies are detailed in Section 3. Conclusions are summarized in Section 4 including a discussion of implications from DNS studies.
\section{Model Description:}
In the Eulerian-Lagrangian particle-based model, a pressure and height dependent parallel FORTRAN algorithm \cite{bhow22} for an upward rising domain is developed and used to investigate the diffusion growth of individual cloud droplets in a hydrostatic turbulent environment. In addition to atmospheric flow velocity (\textbf{U}), temperature ($T$), and mixing ratio ($q_v$), the following prognostic variables are incorporated in macroscopic equations: (i) domain air pressure ($P$) and (ii) domain height ($H$) from the cloud base.

The basic equations for turbulent flow fields used in this article are as follows \cite{bhow22}:
\begin{equation}
    \nabla \cdot {\bf U} = 0 ~~~\label{eq1}
\end{equation}
\begin{equation}
\frac{\partial {\bf U}}{\partial t} + ({\bf U}.\nabla){\bf U} 
= - \frac{1}{\rho_0} \nabla p 
+\nu \nabla^2 {\bf U} + B e_z + f_{LS}(x, t), ~~~\label{eq2}
\end{equation}
\begin{equation}
\frac{\partial {T}}{\partial t} + {\bf U}.\nabla { T} = k \nabla^2 { T}+ \frac{L}{c_p}C_d - \frac{g w_z}{c_p}, ~\label{eq3}
\end{equation}
\begin{equation}
\frac{\partial {{q_v}}}{\partial t} + {\bf U}.\nabla {{q_v}} = D_v \nabla^2 {{q_v}} - C_d. ~\label{eq4}
\end{equation}
Here, $B$ and $f_{LS}$ denote buoyancy term, and forcing term, respectively. $L$ is the latent heat of evaporation, $C_p$ is specific heat, and $C_d$ is the condensation rate. Following \citet{RogYau89}, the condensation rate ($C_d$) is measured in units of mass of condensate per mass of air per unit time. %and defined as \cite{Chen_18}.
The condensation rate inside each grid cell is determined by \cite{bhow22},
\begin{eqnarray}
C_d(X,t) = \frac{1}{m_a} \frac{dm_l(x,t)}{dt} = \frac{4\pi \rho_l K}{\rho_0 {l_\eta}^3}\sum_{\beta = 1}^\Delta S(X_\beta,t)r(X,t)
\end{eqnarray}
Here, $m_a$ is the mass of air per grid cell, the sum collects the droplets inside each grid cell of size ${l_\eta}^3$ that surround the grid point $x$ and $S$ is supersaturation. 

Pressure variation is calculated from the hydrostatic equation:
\begin{equation}
\frac{d{P}}{dt} = -\frac{g {P} w_z}{R_t {{T}}} ~~~\label{eq5}
\end{equation}
Here, $R_t$ is the specific gas constant of moist air and the vector $g = (0,0, -g_z)$ includes the gravitational acceleration $g_z (= 9.81m/s^2)$ and $w_z$  = 0.5 m/s is a constant upward velocity of the domain.

The cloud droplets are modeled as spherical particles and they are smaller than the Kolmogorov scale ($\eta$) throughout the simulation. 
%The motion of cloud droplets suspended in a flow is determined by Stokes’s formula \cite{Max_Ril_83,Elg_Tru_92}, in which the action of fluid is reduced to the sum of a linear viscous drag and of a gravitational force \cite{Nas_Juc_18}. 
The Lagrangian evolution of each cloud droplet is described by \cite{bhow22,Kum14,Pau17,Chen18}:
\begin{equation}
\frac{d{\bf X}(t)}{dt} = {\bf V}({\bf X}, t), ~\label{eq7}
\end{equation}
\begin{equation}
\frac{d{\bf V}({\bf X},t)}{dt} = \frac{{\bf U}({\bf X},t) - {\bf V}({\bf X}, t)}{\tau_p} + g, ~\label{eq8}
\end{equation}
where, \textbf{U(x,t)} is the flow velocity at the droplet’s position \textbf{X(t)}, \textbf{V(X,t)} is the droplet’s velocity, and  $\tau_p$(=$\frac{2\rho_l r^2}{9\rho_0 \nu}$) is the droplet inertial response time \cite{Chen18}. 

The droplet radius $r(X,t)$ is integrated from the diffusional droplet growth equation, which incorporates the solute and curvature effects \cite{bhow22,Kor95},
\begin{equation}
\frac{dr({\bf X},t)}{dt} 
= \frac{K}{r({\bf X},t) + \xi} \left( S({\bf X},t) + 1 - \left(1+\frac{a {r_d}^3}{{r({\bf X},t)}^3 + b {r_d}^3)}\right) exp\left(\frac{B}{r({\bf X},t)}\right)\right), ~\label{eq9}
\end{equation}

Here, $r_d$ is the dry aerosol radius. Following \citet{Kor95}, we considered dry aerosols as sodium chloride (NaCl). $a$ and $b$ are constants and we assume $a=-1.215$ and $b= 1.815$. $K = \frac{D_vE}{\rho_l R_v T (1+K_1)}$, $K_1 = (\frac{L}{R_v T} - 1)\frac{LD_vE}{R_v T^2 k}$, $B = \frac{2\sigma}{\rho_l R_v T}$ , and $\xi = \frac{\frac{D_v}{\alpha \nu} - K_1 \frac{k}{\omega \nu}}{1+K_1}$,
%, and $\nu = (\frac{R_v T}{2\pi})^{1/2}$,
$D_v$ is the vapor diffusivity, $k $ is the thermal conductivity of air, $\alpha$ is the condensation coefficient, $\omega$ is the thermal accommodation coefficient, and $\sigma$ is the coefficient of surface tension of water. More details can be found in Appendix A of \citet{Kor95}.

Supersaturation in each grid cell is calculated as:
\begin{equation}
S = \frac{E}{E_s}-1
\end{equation}
Water vapor pressure ($E(x,t)$) is calculated as:
\begin{equation}
   E = \frac{P q_v M_d}{q_v M_d + M_w}
\end{equation}
and the saturation water vapor pressure ($E_s(x,t)$) is calculated from the Clapeyron equation.

It is important to mention here that the values of variables- latent heat of evaporation ($L$), the thermal conductivity of air ($k$), specific heat ($c_p$), the diffusivity of water vapor ($D_v$), $R_t$, and all other related variables change with corresponding pressure, temperature and height of the upward rising domain. For convenience of reference, a list of constants is given in Table \ref{tab-1}).

To discretize the partial differential equations, the high-resolution pseudo-spectral method using Fast Fourier Transformations (FFT) is adopted \cite{Pop00,FriJoh05,Pek08,Kum12}. All the time, height and pressure dependent ordinary equations are solved numerically by the second order predictor-corrector method.
\begin{table} [h!]
    \centering
    \begin{tabular}{ || c c c c || }
    \hline
    \textbf{Parameters}  & \textbf{Symbols} & \textbf{Units} & \textbf{Values} \\ [0.5 ex]
    \hline \hline
 Thermal capacitance of   &  &  &  \\
 dry air under constant pressure & $C_{pa}$ & J/Kg K & 1.005 $\times 10^3$ \\
Thermal capacitance of  & &  &  \\
water under constant pressure & $C\_{pv}$ & J/Kg K & 1.850 $\times 10^3$ \\
Weight of dry air & $M\_a$ & Kg & 1 \\ 
Molecular Weight of dry air & $M\_d$ & Kg & 0.02896 \\ 
Molecular Weight of water & $M\_w$ & Kg & 0.01806 \\ 
Universal Gas Constant & $R\_g$ & J/kg K  &	8.317 \\ 
Gas Constant for water vapor & $R\_v$ & J/kg	& 461.5 \\ 
Kinematic viscosity	& $\nu$ & $m^2/s$ & $1.5 \times 10^{-5}$ \\ 
Reference mass density of air & $\rho\_0$ & $kg/m^3$ & 1.06 \\ 
Weight Density of dry aerosol & $\rho\_d$ & $kg/m^3$ & $2.5 \times 10^3$ \\
Weight Density of Water& $\rho\_l$ & $kg/m^3$ & $10^3$ \\ 
Dissipation rate & $\epsilon\_d$ & $cm^2/s^3$ & 33.75 \\ 
Taylor microscale Reynolds number & ${Re}_\lambda$& & 59\\ \hline 
    \end{tabular}
    \caption{List of constants}
    \label {tab-1}
\end{table}
\subsection{Initial Setup $\&$ design of experiments}
\begin{figure}[h!]
\centerline{\includegraphics[height=7cm,width=9cm]{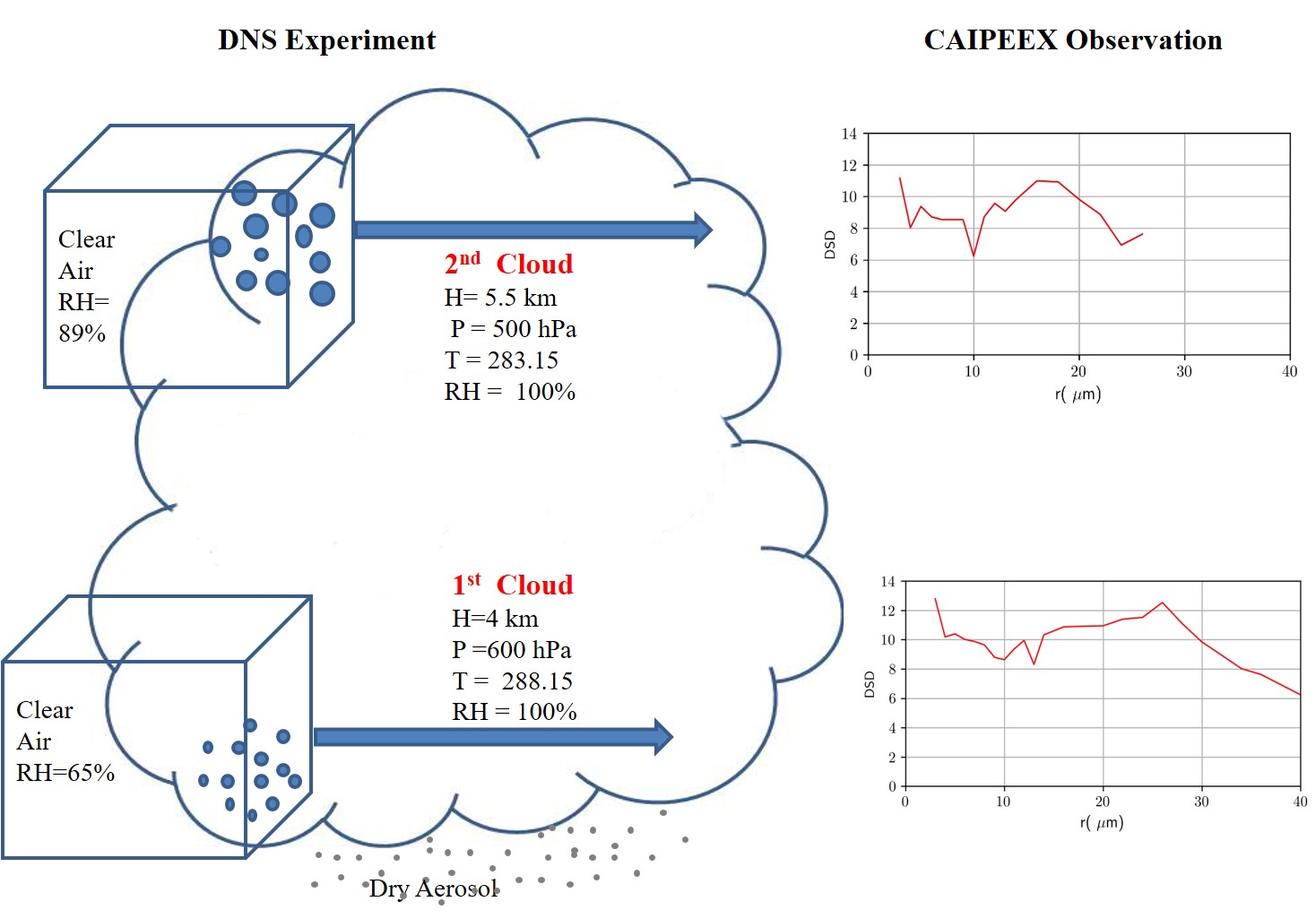}}
\caption{{Schematic diagram of the DNS experiment at two maximum levels the parcels attained through ascent. 
%Cloud droplet size distribution at two altitudes (2.5 km and 4 km) of 1 Hz resolution used in the DNS are shown on the panel. 
These two cloud passes were considered from CAIPEEX observation over Solapur, India (Lon: 75.84 E and Lat: 17.72 N) on 23 August 2019. The cloud droplet size distributions in the diameter range 2-50 $\mu m$ were measured by Cloud Droplet Probe manufactured by Droplet Measurement Technologies, USA.}}
\label{fig-1}
\end{figure} 
Direct numerical simulations (DNS) are performed for a domain of size 25.6 cm $\times$ 25.6 cm $\times$ 25.6 cm with 256 $\times$ 256 $\times$ 256 grid points. Thus, a 1 mm grid resolution is applied. To initiate the clear-cloudy mixing process, a saturated cloud slab of size 10 cm $\times$ 25.6 cm $\times$ 25.6 cm is placed in the domain's center. Two types of cloud passes are taken from the CAIPEEX experiments (Figure \ref{fig-1}) when setting up the DNS studies. 
\begin{table} [h!]
    \centering
    \begin{tabular}{ || c c c || }
    \hline
      & \textbf{1$^{st}$ cloud} & \textbf{2$^{nd}$ cloud} \\ [0.5 ex]
    \hline \hline
 Droplets $\&$ dry aerosols number Concentration ($/cc$) & 48.126	 & 11.413 \\
Initial Temperature (K)	& 288.15	& 283.15 \\ 
Initial Pressure (hPa) 	& 600	& 500 \\ 
Initial Height of the Domain (Km) & 4	& 5.5 \\ 
Initial Relative Humidity (RH) & 65$\%$ &	89$\%$ \\
Mixing of Clear-Cloudy Air	& Present &	Present \\ \hline 
    \end{tabular}
    \caption{Initial Setup of DNS experiment guided by the observation, CAIPEEX.}
    \label {tab-2}
\end{table}
\par
The simulation set ups at cloud base (1$^{st}$ cloud, less RH) and inside or top (2$^{nd}$ cloud, more RH) are presented in Table \ref{tab-2}. At the beginning of the simulations, the cloud droplets are placed randomly inside the cloud slab and allowed to move freely with turbulent air. The simulation domain is treated as adiabatic, which means that once the cloud droplets and their corresponding dry aerosols are placed within it at the beginning of the simulation, they will remain there throughout the simulation.
\par
There is no activation of dry aerosols, no sedimentation, or collision-coalescence processes inside the domain. This means the growth of the droplets is driven only by condensation/evaporation. At every time step of the foregoing simulations, droplets with diameters (D) less than 0.1 $\mu m$  are considered evaporated completely and removed from the system.  
%we have loosened the requirement that cloud droplets have crucial diameter (D) greater than 10 $\mu m$ (except when computing DSD) based on CAIPEEX observation. 
All the experiments are carried out with seven different dry aerosol sizes (see Table \ref{tab-3}).
\begin{table} [h!]
    \centering
    \begin{tabular}{ || c c c c c c c c || }
    \hline
%      & \textbf{1$^{st}$ cloud} & \textbf{2$^{nd}$ cloud} \\ [0.5 ex]
 %   \hline \hline
 Dry aerosol radius & $r_d{_1}$ & $r_d{_2}$ & $r_d{_3}$ & $r_d{_4}$ & $r_d{_5}$ & $r_d{_6}$ & $r_d{_7}$   \\ \hline
 $\mu m$ & 0.05 & 0.07 & 0.09 & 0.1 & 0.5 & 0.7 & 0.9\\ \hline 
    \end{tabular}
    \caption{Dry aerosols’ radii used for the simulations are obtained from CAIPEEX observation.}
    \label {tab-3}
\end{table}
\section{Results}
The growth of a cloud droplet is initially governed by the diffusion of the water vapour molecules towards the droplet. There are simultaneous condensing and evaporating of water vapour at the surface of a droplet. Therefore, two important phenomena that influence the growth by diffusion are the curvature effect and the solution effect. In this present work, we primarily focused on the curvature effect by considering different dry aerosol sizes.  The initial size of dry aerosols is important for activation of the droplets based on K$\ddot{o}$hler equation \cite{PruKle10}. The hygroscopicity also depends on the size of dry aerosols. It is important to note that the aerosols with dry diameter larger than few micrometers, which is called giant cloud condensation nuclei (GCCN) provides rain embryo and can broaden the size distribution \cite{Joh82,Fei99}. The curvature of a droplet tends to increase the concentration of vapour at the surface of the droplet. Since diffusion is the movement from higher concentrations to lower concentrations, the curvature effect tends to delay droplet growth by diffusion. As a droplet grows, its curvature decreases and becomes more like a plane surface and the influence of the curvature effect decreases as well. 
\par
\begin{figure}[htbp]
\centerline{\includegraphics[height=8cm,width=10cm]{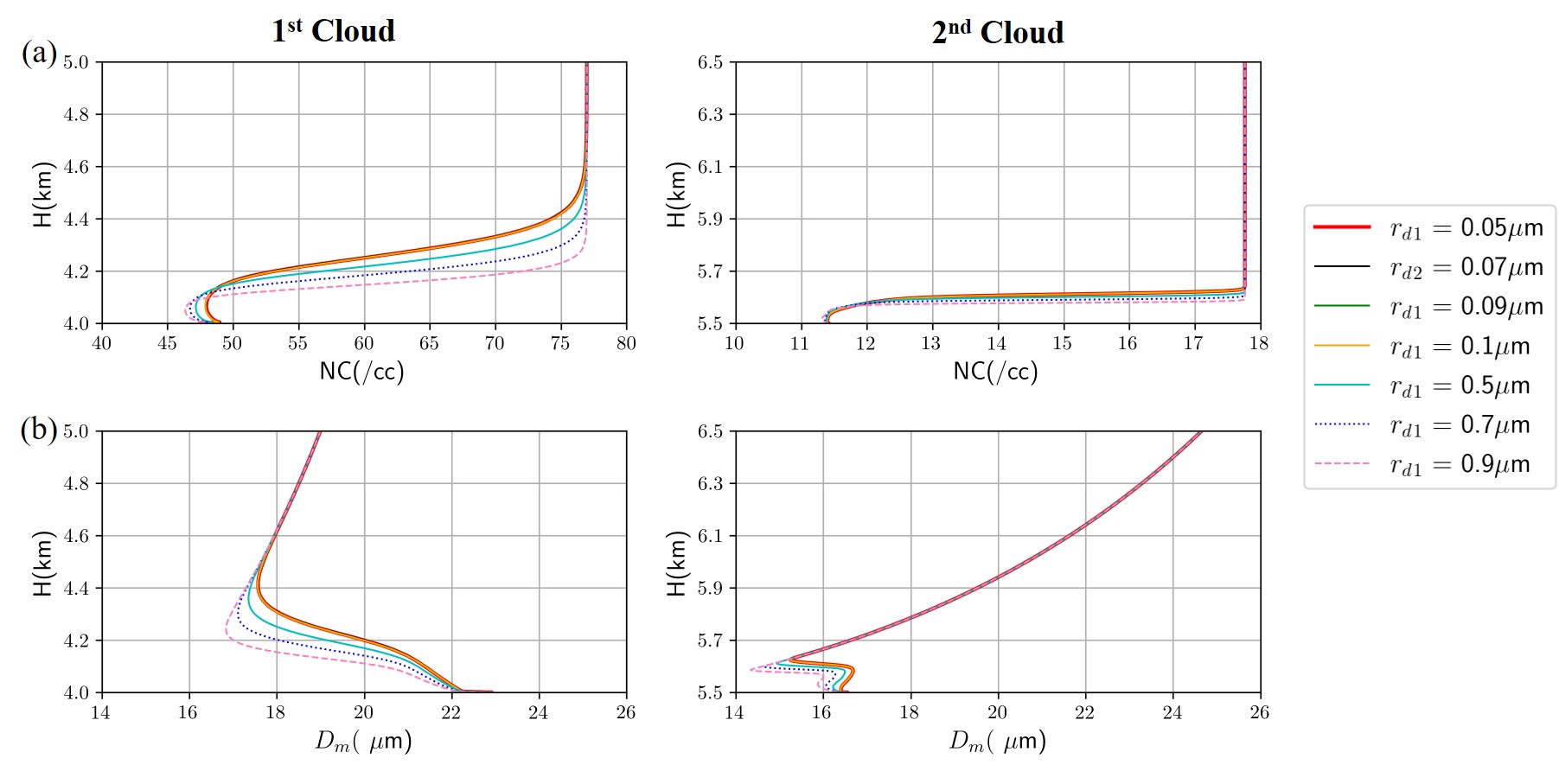}}
\caption{Changes in cloud droplets (a) number concentrations and (b) mean diameter ($
D_m$), inside the rising domain with respect to height for different sizes of dry aerosols in less humid (1st cloud) and more humid (2nd cloud) conditions.}
\label{fig-2}
\end{figure}
\begin{figure}[htbp]
\centerline{\includegraphics[height=9cm,width=10cm]{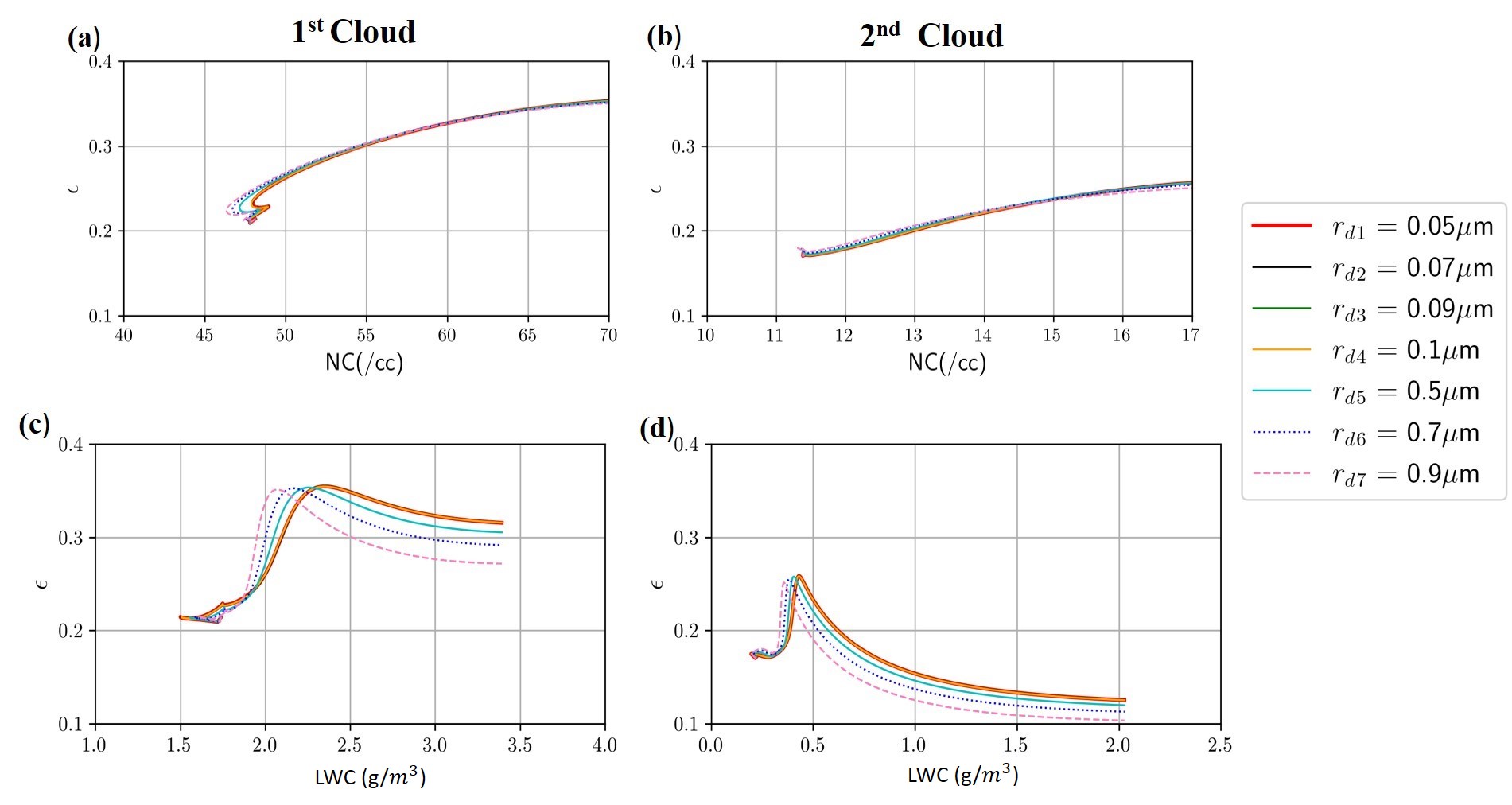}}
\caption{The relationship between relative dispersion ($\epsilon$) and cloud droplet number concentration (NC) (a,b) and the variation of dispersion with liquid water content (LWC) for two types of clouds (c,d) over Indian subcontinent.}.
\label{fig-8}
\end{figure}

The vertical profiles of temperature, pressure and domain’s upward velocity obtained from DNS experiments demonstrate (supplementary Figure S1 and Figure S2) the differences between the two types of clouds (1$^{st}$ cloud – less humid condition, 2$^{nd}$ cloud – more humid condition). The low temperature and pressure at the cloud top (2$^{nd}$ cloud) as compared to the cloud base (1$^{st}$ cloud) indicates the fidelity of DNS experiments. Now, it is important to investigate the growth of cloud and rain droplets in the two cloud systems or environments. 
Figure \ref{fig-2} shows the variation in cloud droplets number concentration (NC) and droplets mean diameter ($D_m$) with height in an upward rising domain. It is important to note that, based on CAIPEEX observations \cite{Kon12} we consider the droplets $\ge ~10 ~\mu m $  in order to calculate NC and $D_m$. The 1$^{st}$ cloud is drier (less RH) as compared to 2$^{nd}$ cloud (which is moist, higher RH) and as a result more (less) droplet number concentrations are observed (Figure \ref{fig-2}a) in 1$^{st}$ cloud (2$^{nd}$ cloud). Near the cloud base there is a decrease in the number concentration of droplets larger than 10 $\mu m$ diameters (Figure \ref{fig-2}a). This implies that in-homogeneous entrainment mixing is the dominant process at the cloud boundaries, similar to the findings of \citet{Sma13}. On the other hand, the change in number concentration and mean diameter of cloud droplets ($\ge ~10 ~\mu m $) is insignificant at the cloud top. Interestingly, mean droplet diameter increases (decreases) at the cloud top (cloud base) as seen in Figure \ref{fig-2}b. The larger (smaller) mean size of cloud condensate in more humid (less humid) cloud from DNS (Figure \ref{fig-2}b) is revealed in this study, which is similar to observation and NWP study \cite{Haz13}. 
\begin{figure}[htbp]
\centerline{\includegraphics[height=9cm,width=10cm]{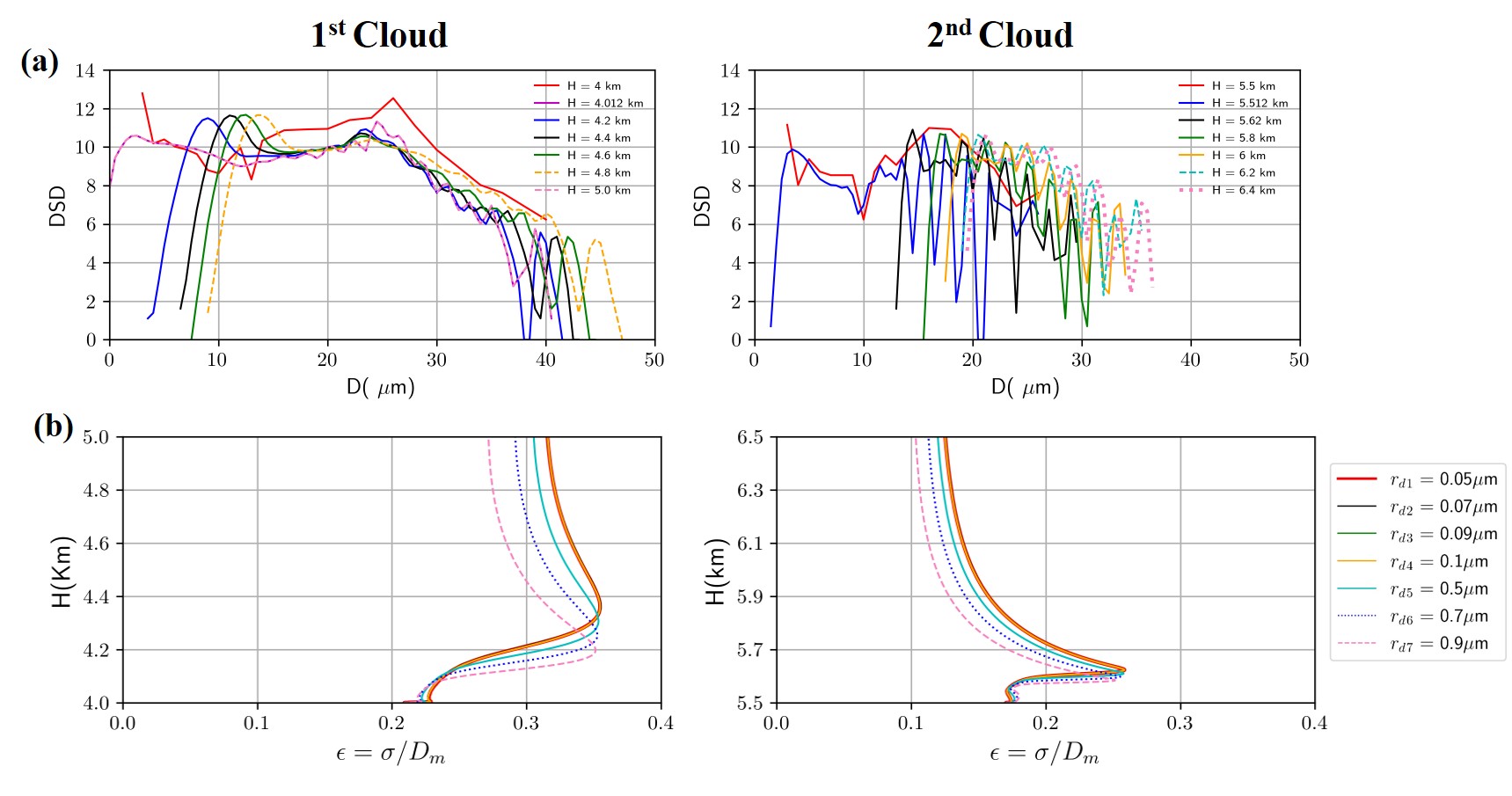}}
\caption{(a) The droplet size distribution (in log scale) at different altitude for $r_{d_3}$ and (b) relative dispersion value ($\epsilon  = \sigma /D_m$) with respect to height for seven different sizes of dry aerosols in less humid (1st cloud) and more humid (2nd cloud) conditions. }.
\label{fig-3}
\end{figure} 
Figure \ref{fig-3}a shows the droplet size distribution (DSD) with height  for the two types of cloud systems (1$^{st}$ cloud with less liquid water content (LWC) and 2$^{nd}$ cloud with more LWC). To calculate DSD, the entire range ($D > 0.1 \mu m$) of cloud droplets' presence inside the domain is divided into a set number of bins.  The DSD was then computed based on the number of droplets that lie in each bin and plotted on log scale. Cloud droplet size distribution at different heights demonstrate that shallow cloud (i.e., 1$^{st}$ cloud) disperse more (Figure \ref{fig-3}a) as compared to cloudy system (i.e., 2$^{nd}$ cloud). The increases in dry aerosol size enhance the surface area of the substrate and produce more droplets, particularly at the cloud base (Figure \ref{fig-2}a), which is not significant at the top or inside the cloud. At the cloud base, the critical size of activation as determined by K$\ddot{o}$hler theory (curvature effect) is  crucial \cite{PruKle10,Che94,CheLiu04,CheWanChe07}. Therefore, with these DNS tests, we further found that the cloud drop number concentration is rather sensitive to this activation size because it influences the overall diffusional growth \cite{PruKle10,Che94,CheLiu04,CheWanChe07}. Thus, the dry aerosol radii are an important parameter to determine the number concentration of smaller droplets through the curvature effect (Kelvin equation) and also impact on super-saturation via evaporation and condensation processes \cite{PruKle10}. The warm-cloud processes considered are cloud drop activation, drop growth by vapor diffusion, collision-coalescence, and break-up \cite{CheLiu04,CheWanChe07}. The theories of cloud drop activation and subsequent growth by vapour diffusion are well established \cite{PruKle10}. The more cloud drop number concentrations with lesser size are activated \cite{Two59} due to an increase in dry aerosol size (Figure \ref{fig-2}a) at the cloud base. We have also calculated relative dispersion value ($\epsilon  = \sigma /D_m$), from both clouds, where $\sigma$ is the standard deviation and $D_m$ is the mean diameter. The relative dispersion value is not very sensitive to the vertical height above the cloud base (Figure \ref{fig-3}b).  On the other hand, the width of the droplet size distributions decreased from cloud base to cloud top (Figure \ref{fig-3}).  The results are very similar to observation of Flight data measured in warm convective clouds near Istanbul in June 2008 \cite{Tas15}. 
Overall, the values of relative dispersion ($\epsilon$) vary in the range of 0.2 (0.1) to 0.37 (0.26) for 1$^{st}$ cloud (2$^{nd}$ cloud). Note that for undiluted clouds (nearly adiabatic), the relative dispersion is about 0.36, which is a typical value for cumulus cloud \cite{Pol93,KhaPin18,KonPra21}. This is in agreement with previous studies which indicated that $\epsilon$ tends to be bounded in a similar narrow range in warm cumuli \cite{Chen18,Pan12,Ber11}, stratus clouds \cite{Peng07} and stratocumulus clouds \cite{Paw06}. The relative dispersion ($\epsilon$) is clearly seen as a decreasing due to increase of LWC.
Figure \ref{fig-8}(a,b) represents the relation of relative dispersion and number concentrations of cloud droplets. There is a linear relationship between relative dispersion ($\epsilon$) and cloud droplet number concentration (NC) as revealed from this study using CAIPEEX observation over Indian subcontinent. \citet{MarJohSpi94} have also shown the similar formulation where the cloud droplet spectral dispersion is a linear function of cloud droplet number concentrations. This study also shown that the relative dispersion ($\epsilon$) is also varying with liquid water content (Figure \ref{fig-8}(c,d)), which is dependent on type of clouds and dry aerosol size.
\section{Discussions, Conclusions $\&$ Implications:}
\begin{figure}[htbp]
\centerline{\includegraphics[height=5cm,width=10cm]{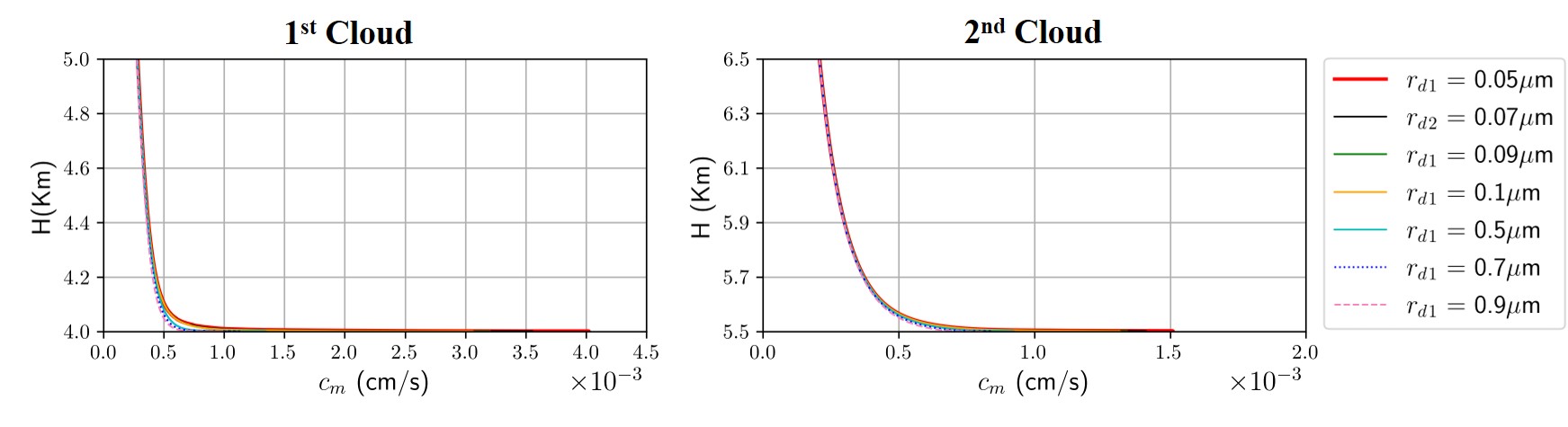}}
\caption{The coefficient in the diffusional growth rate equation is calculated from two clouds (1st cloud, less humid and 2nd cloud, more humid).}.
\label{fig-7}
\end{figure} 
The importance of diffusional growth of droplets in real environment can be varied from shallow cloud to deep convective cloud system due to variation of supersaturation and aerosol number concentrations. Therefore, in the calculation of growth rate ($dr/dt$) of droplets (Eq. \ref{eq9}), the proper choice of $c_m$ (i.e, $\frac{K}{1 +\xi/r(X,t)}$) is very important in model to reduce the underestimation or overestimation of smaller and larger droplets (Figure \ref{fig-7}). The selection of coefficient ultimately can help to simulate light, moderate and heavy rainfall in numerical model, which are presented in a companion paper (Part-II of \cite{Bhow23b}) of the same journal. 
\par
The important conclusion from this DNS study is that the diffusional growth is very sensitive to supersaturation, aerosol number concentrations and dry size of aerosols. Many of the key properties of an observation may be captured by DNS, which is crucial for numerical modelling. Since the accuracy of the parameterization scheme is limited by the preciseness of the dataset, it is essential to use a DNS study that can provide deeper understanding of the droplet growth processes and the evolution of cloud particle size spectra under a broad range of conditions. 
%In this present study, we have relaxed the criterion for cloud drops to be of size greater than either 10 $\mu m$ as their critical size based on CAIPEEX observation. 
Furthermore, some of the largest cloud condensation nuclei (CCN) are activated directly into raindrops, which are called the rain embryos. It is also concluded that in the shallow cloud, the relative dispersion is larger than that in a  cumulus cloud and therefore, shallow clouds precipitation development is not only determined by the liquid water content but also the available time for rain initiation. Although we have considered the curvature effect (i.e., Kelvin effect), but in this present equation, we neglect the effects of hygroscopicity of aerosols. Therefore, in future, we will consider particles for which the “chemistry” terms or solute effect of aerosols in the diffusional growth process. The results highlight the importance of model development by incorporating new coefficient of diffusional growth process and cloud droplets spectral dispersion over Indian summer monsoon clouds. 
\par
%In the earlier studies, it has been shown that the modifications of 'collision-coalescence' and ‘auto-conversion’ processes in climate model are able to improve heavy rainfall category but unable to progress in light rainfall bins \cite{Mud_22, Dut_Haz_21}. The development of rainfall in light rainfall categories is not explored so far. Therefore, this is first time, we have attempted to demonstrate the pathway to reduce the overestimation of lighter rain through “tuning” the diffusional growth rate coefficients guided by DNS.
%
%\begin{figure}[htbp]
%\centerline{\includegraphics[height=9cm,width=10cm]{Fig-11.jpg}}
%\caption{The relationship between relative dispersion ($\epsilon$) and cloud droplet number concentration (NC) (a,b) and the variation of dispersion with liquid water content (LWC) for two types of clouds (c,d) over Indian subcontinent.}.
%\label{fig-8}
%\end{figure} 
%
\par
As mentioned earlier, the Numerical Weather Prediction (NWP) model required better “autoconversion” parameterization for the precipitation calculation. \citet{WuXiDonZha18} have shown that the “autoconversion” initializes the formation of precipitation. It is also responsible for the formation of drizzle in stratiform clouds as suggested by \citet{LiuDau04}. Different “autoconversion” parameterization schemes are suggested by past researchers (e.g., \citet{Beh94,BerRei74,KhaKog00,TriCot80} and these parameterization schemes can be classified according to the choice of parameters such as, cloud water content, droplet number concentration, and relative dispersion of cloud droplets \cite{XieLiu15}. \citet{Kes69} proposed a simple autoconversion parameterization as below, where $K$ is the autoconversion coefficient, LWC and $L_c$ are the cloud water content and its threshold value, respectively \cite{GhoJon98}:
\begin{equation}
    \frac{dLWC}{dt} = K (LWC -L_c)
\end{equation}
\citet{Kes69} assumed that the precipitation rate is directly proportional to the cloud water content. Later, \citet{Sun78} proposed an alternative type of autoconversion parameterization, which is known as Sundqvist-type parameterization, namely, 
\begin{equation}
    P_s = c_s LWC \left(1- exp \left[-\left(\frac{LWC}{L_c}\right)\right]\right) ~\label{eq59}
\end{equation}
Here, $c_S$ is an empirical constant in $s^{-1}$. 

It is worth stressing that the primary difference bewteen Kesser-type and Sundqvist-type autoconverstion parameterization lies in the treatment of the threshold value of cloud liquid water content (i.e. the value of $L_c$).

%
%\citeA{KhaKog00} scheme mentioned that autoconversion rate decreases with the increasing of cloud droplet number concentrations and increases with increasing liquid water content. 
It should be noted that this autoconversion parameterization scheme is very commonly used in general circulation models \cite{Loh07,MorGet08}. The coupled climate model (CFSv2) is presently for the operational seasonal forecasting in India \cite{Rao19}. The Sundqvist type autoconversion scheme \cite{Sun89} is used in the microphysical scheme of CFSv2 \cite{Haz17}. The relative dispersion based autoconversion scheme is proposed by \citet{LiuDau04}, which assumes that the autoconversion rate is related to the cloud water content (LWC), droplet number concentration (NC), and relative dispersion ($\epsilon$) of cloud droplets. \citet{LiuDauMc04} generalized the Sundqvist type parameterizations and demonstrated the improvement considering the relative dispersion, cloud liquid water content, droplet concentration, more realistically than that of conventional Kessler type. $P_D$ ($gm/{cm}^3 {s}^{-1}$) is the cloud-to-rain autoconversion rate:
\begin{equation}
P_D = a_1 \left[\frac{(1+3\epsilon^2)(1+4\epsilon^2)(1+5\epsilon^2)}{(1+\epsilon^2)(1+2\epsilon^2)} \cdot {N_c}^{-1}{LWC}^3\right]\beta, ~\label{eq19}
\end{equation}
\begin{equation}
\beta = 0.5({x_c}^2+2x_c+2)(1+x_c) ~exp\left(-2x_c\right)
\end{equation}
%
%Where, $NC$ and $LWC$ are the cloud droplet number concentration (${cm}^{−3}$) and cloud water content ($gm/{cm}^{3}$) respectively.
\par
where, $x_c$ is an analytic parameter and defined as,
\begin{equation}
    x_c = a_2{NC}^{3/2} {LWC}^{-2}
\end{equation}
$a_1$ , $a_2$ are constants and values are $1.1\times{10}^{10}$and $9.7\times {10}^7$, respectively.
\begin{figure}[htbp]
\centerline{\includegraphics[height=9cm,width=15cm]{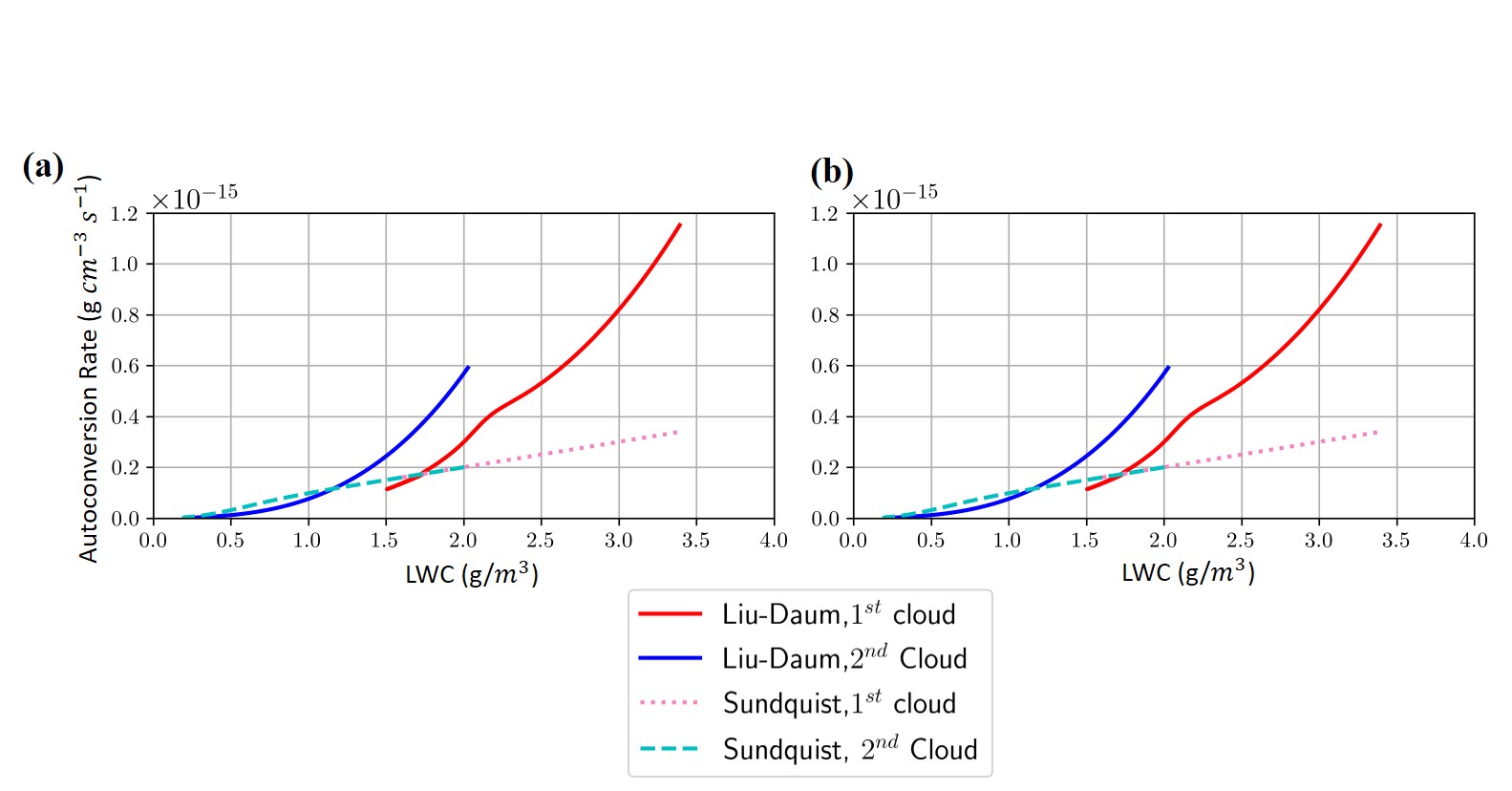}}
\caption{Comparison of Autoconversion rate for (a) $r_{d_3}$ and (b) $r_{d_4}$ calculated using Liu-Daum type parameterization and Sundquist typeparameterization for two types of ISM clouds.}
\label{fig-99}
\end{figure} 
To verify the points mentioned in \citet{LiuDau04}, the autoconversion rate using sundqvist-type and Liu-Daum -type parameterization schemes as a function of LWC for both clouds are shown in Figure \ref{fig-99}(a,b). The primary difference between these two parameterization schemes can clearly been observed from the figure. Moreover, it is evident from both figure \ref{fig-99}(a) and figure \ref{fig-99}(b) that the Liu-Daum Kessler-type autoconversion rate for 1$^{st}$ (less humid) cloud and 2$^{nd}$ (more humid) cloud are following different trends, which seems to be more realistic. On the other hand, not much variation is observed in the Sundqvist-type autoconvresion rate for both clouds.

%\par
%Figure\ref{fig-8}(a,b) represents the relation of relative dispersion and number concentrations of cloud droplets. There is a linear relationship between relative dispersion ($\epsilon$) and cloud droplet number concentration (NC) as revealed from this study using CAIPEEX observation over Indian subcontinent. \citeA{MarJohSpi94} have also shown the similar formulation where the cloud droplet spectral dispersion is a linear function of cloud droplet number concentrations. This study also shown that relative dispersion ($\epsilon$) is also varying with liquid water content (Figure\ref{fig-8}(c,d)), which is dependent on type of clouds and dry aerosol size. 

\par
\begin{figure}[htbp]
\centerline{\includegraphics[height=9cm,width=10cm]{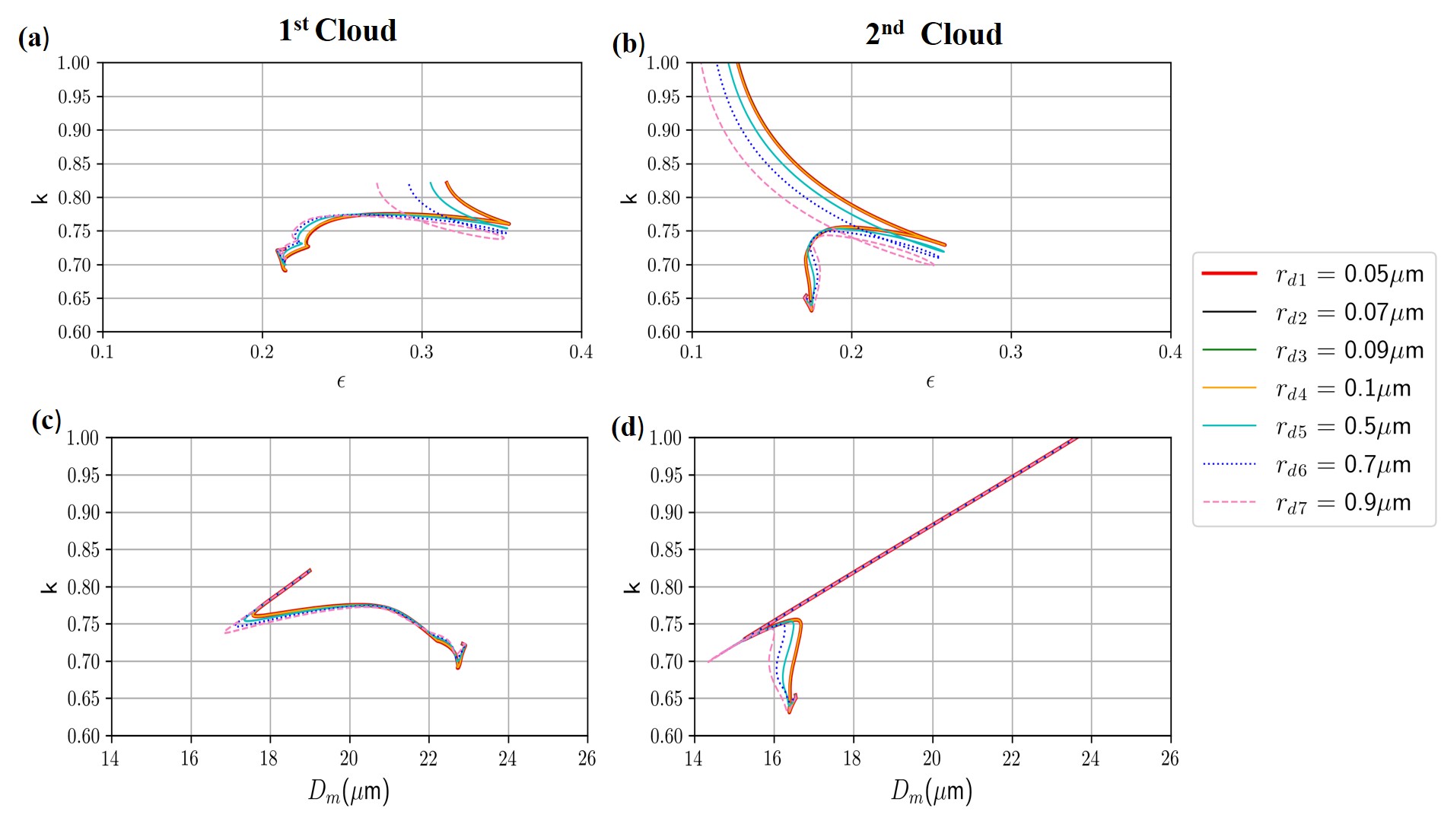}}
\caption{The gradient (k) varies with relative dispersion (a,b) and mean volume diameter of cloud droplet (c,d) for two types of clouds (c,d) over Indian subcontinent.}.
\label{fig-9}
\end{figure} 

\par
It is also noted that radiation schemes used in large-scale numerical models are very sensitive to the effective radius \cite{RanCoa84}. There is a linear relationship between effective radius and mean volume radius with smaller gradient (k). \citet{Sli89} recommended that the dependence of cloud radiative properties on the liquid water path and effective radius. Therefore, the understanding of k and its relation with mean diameter and dispersion will be valuable information for the parameterization of effective radius for cloud-climate feedbacks \cite{MarJohSpi94,CheWanChe07}. The suggested parameterization based on aircraft observation proposed by \citet{MarJohSpi94} are also revisited using DNS studies. Figure \ref{fig-9}(a,b) depicts the relationship between the gradient (k) and dispersion. Similarly, the gradient (k) and mean volume radius connection is also presented in Figure \ref{fig-9}(c,d). It is interesting to note that gradient (k) varies significantly from shallow to convective clouds. The value of k is $\sim$ 0.75 at the relative dispersion of 0.2 but it can varies from 0.70 - 0.82 (1$^{st}$ cloud) and 0.65 - 1.0 (2$^{nd}$ cloud). The value of gradient (k) is very similar (0.67 - 0.88) to earlier studies by \citet{MarJohSpi94} from air bore observation.  
\begin{equation}
    r_e = \left(\frac{3LWC}{4 \pi \rho_w k ~NC}\right)^{1/3} ~\label{eq20}
\end{equation}
Therefore, the newly designed dispersion based “autoconversion” scheme for Indian region would be useful for the Indian summer monsoon precipitation calculation in general circulation model and can be adopted for the model development activity. We have implemented dispersion based “autoconversion” in climate forecast system (CFS) and modified the diffusional growth rate of cloud droplets. The interesting results from CFS studies are presented in another paper (Part-II of \cite{Bhow23b}).
\par
It is also concluded that parameterization of effective radius, which is important for cloud-radiation feedback may be modified in climate model guided by air borne observation and DNS results as mentioned in this study. 
\section{Acknowledgements}
The IITM Pune is funded by the Ministry of Earth Science (MoES), Government of India. The authors acknowledge the use of high-performance computational resources at IITM, particularly "Pratyush" HPC system without which this work would not have been possible.  Special thanks to Dr. Mahin Konwar for providing us CAIPEEX data and for his insightful comments in shaping the first draft of this paper.  Authors thank CAIPEEX team members for their sincere efforts in successfully conducting airborne observations.
\end{document}